\documentclass[aps,pra,twocolumn,showpacs]{revtex4}
\usepackage{graphicx} 
\usepackage{amsmath}
 
\newcommand{\beq}{\begin{equation}}
\newcommand{\enq}{\end{equation}}
\newcommand{\bea}{\begin{eqnarray}}
\newcommand{\ena}{\end{eqnarray}}
\newcommand{\la}{\langle}
\newcommand{\ra}{\rangle}

\begin{document}

\title{Quantum states of $p$-band bosons in optical lattices}
\author{A. Collin}
\author{J. Larson}
\author{J.-P. Martikainen}
\affiliation{NORDITA, 106 91 Stockholm, Sweden}
\date{\today}

\begin{abstract}
We study a gas of repulsively interacting 
bosons in the first excited band of an optical lattice.
We explore this $p$-band physics both within the framework
of a standard mean-field theory as well as with the more accurate
generalized Gutzwiller
ansatz. We find the phase diagrams for two- and three-dimensional
systems and characterize the first Mott-states in detail.
Furthermore, we find that even though the $p$-band 
model has strongly anisotropic
kinetic energies and inter-flavor interaction terms are missing
in the lowest band theory,
the mean-field theory becomes useful
quite rapidly once the transition from the Mott-insulator
to the superfluid is crossed.

\end{abstract}

\maketitle 

\section{Introduction}
\label{sec:intro}
Systems of cold atoms in optical lattices have seen a
dramatic experimental progress in the recent past~\cite{Bloch2008a,Lewenstein2007a}.
Due to realization of optical lattices, low densities, and low temperatures, 
a fantastic degree of control has been obtained which has made detailed studies 
of strongly correlated quantum systems possible. For example,
the Mott-superfluid transition~\cite{Jaksch1998a,Greiner2002a} 
has been successfully observed in optical lattices.  This transition, due to the competition between kinetic energy and repulsive on-site interactions between lowest band bosons,
can occur even at $T=0$ and is therefore driven by quantum fluctuations. For large 
interactions, the energy is minimized in an incompressible state with fixed
atom numbers at each lattice site, while for weaker interactions the kinetic energy favors
atomic tunneling which drives the system into a superfluid.

The early experiments were confined to the lowest band and while 
increasing interactions can make excited band populations 
non-negligible~\cite{Kohl2006a}, the lowest band still dominates. In fact, for very strong interactions, it has been theoretically shown that the lowest band Mott insulator turns into a Mott insulator at the $p$-band~\cite{Alon2005a}. Experimentally, however, atomic population residing on the excited bands is obtained by couple atoms from the lowest band to the excited bands. 
This was experimentally demonstrated by accelerating the lattice for a short period~\cite{Browaeys2005a}, or more recently by coupling atoms from the lowest band Mott insulator into the first excited $p$-band of the lattice via Raman transitions between bands~\cite{Mueller2007a}. In the latter of these two, it was in particular found that the lifetimes of $p$-band atoms are considerably 
longer than the tunneling time-scale in the lattice and they were also able to 
explore how coherence on the excited band establishes. These experiments pave the way to explore also  equilibrium physics of the purely $p$-band bosons~\cite{Isacsson2005a} and furthermore provide possible routes to realize supersolids~\cite{Scarola2005a} or novel phases~\cite{Liu2006a,Xu2007a} on the excited 
bands of an optical lattice. An alternative way to populate higher bands is by considering fermions with a filling factor larger than one~\cite{Wu2008a,CongjunWu2008a,Martikainen2008a}. In this case the Pauli exclusion principle ensures that the fermions that cannot populate the lowest band, must 
occupy the excited bands~\cite{Kohl2005a}.  

In this paper we explore the properties of bosons occupying the first excited bands of an optical lattice. In a periodic potential where the lattice depths are equal in all directions, the (non-interacting) bands are degenerate and a multi-flavor description of the quantum states of atoms is required~\cite{Isacsson2005a,Baillie2009a}. This fact together with the non-isotropic tunneling on the $p$-band add new features and possibilities both for the description of the superfluid
as well as insulating phases. For example, onsite flavor changing collisions can induce
fluctuations in the number of atoms of different flavors even in the insulating phases, giving them non-trivial characteristics. Furthermore, such collisions together with anisotropic tunneling cause different types of phaselockings (both locally as well as between sites) between flavor condensates in the broken symmetry phases.

While in many places we confirm the general picture provided by the somewhat simplified model considered by Isacsson and Girvin~\cite{Isacsson2005a}. Nonetheless, we also find differences which arise due to; the use of real Wannier functions (as opposed to the approximated ones given by a harmonic ansatz), through the inclusion of nearest neighbor tunneling in all directions, or through difference in accounting for the inter-flavor interactions. For future reference, we also compare the Gross-Pitaevskii type mean-field theory with the more accurate Gutzwiller approach and find the parameter regions where the Gross-Pitaevskii approach is reasonably accurate.

The paper is organized as follows. In Sec.~\ref{sec:theory} we derive our model Hamiltonian
and by taking anharmonicity of the lattice potential into account, we outline under what conditions the physical description can be restricted to the first excited $p$-band. We then proceed by deriving  mean-field Gross-Pitaevskii equations for the $p$-band bosons and discuss salient features of their solutions for a homogeneous system both for two-dimensional as well as for three-dimensional systems in 
Sec.~\ref{sec:GPsolutions}. In Secs.~\ref{sec:Gutzwiller2D} and ~\ref{sec:Gutzwiller3D}, we study the $p$-band physics in two- and three-dimensional systems employing the Gutzwiller ansatz and outline the ways how these solutions differ from the mean-field ones. We conclude with a brief discussion in Sec.~\ref{sec:conclusions}.

\section{Formalism}
\label{sec:theory}
The microscopic Hamiltonian for the dilute Bose gas
at low temperatures in a trap is given by
\begin{equation}
\begin{array}{lll}
\hat{H}_{micro} & = & \displaystyle{\int d{\bf r}\hat\psi^\dagger({\bf r})\left[
-\frac{\hbar^2\nabla^2}{2m}+V({\bf r})
\right]\hat\psi({\bf r})}\\ \\
& & \displaystyle{\!+\!\frac{g}{2}
\hat\psi^\dagger({\bf r})\hat\psi^\dagger({\bf r})
\hat\psi({\bf r})\hat\psi({\bf r})
\!-\!\mu\hat\psi^\dagger({\bf r})\hat\psi({\bf r})},
\end{array}
\label{eq:Hmicroscopic}
\end{equation}
where $\mu$ is the chemical potential, $m$ the atomic mass, $g$ is the interatomic interaction strength, and  $\hat\psi({\bf r})$ and $\hat\psi^\dagger({\bf r})$ are the bosonic annihilation and creation operators respectively, while $V({\bf r})$ is the external trapping potential which in this
work is taken to be a lattice potential
\beq
V({\bf r})=V_L\sum_{\alpha\in \{x,y,z\}} \sin^2\left(
\frac{\pi {\bf r}_\alpha}{d}
\right),
\enq
where $d$ is the lattice spacing and $V_L$ the lattice depth. For a deep lattice it is reasonable  to expand the field operators in terms of the localized Wannier functions. Here we go beyond the
usual lowest band Hubbard model by also including the
first excited states ($p$-band). In a three dimensional lattice
this implies an expansion of the field operators
\beq
\hat\psi({\bf r})=\sum_{{\bf i},\sigma} w_{\sigma,{\bf i}}({\bf r}) 
\hat\psi_{\sigma,{\bf i}},
\enq
where ${\bf i}=(i_x,i_y,i_z)$ labels the lattice site and 
$\sigma\in \{0,x,y,z\}$ is the flavor index. 
The bosonic operators $\hat\psi_{\sigma,{\bf i}}$ annihilate
a boson of flavor $\sigma$ from the site ${\bf i}$.
We compute the Wannier functions  from the ideal gas
band structure calculations. 
In this paper 
we assume that the system has been prepared on an excited $p$-bands
and in the following
set the population of the lowest band to zero.

Substituting the operator expansions into the 
Eq.~(\ref{eq:Hmicroscopic}) and ignoring all but the leading
order onsite interactions and nearest neighbor tunneling
processes we derive our fundamental Hamiltonian
\beq
\hat{H}=\hat{H}_0+\hat{H}_{nn}+\hat{H}_{FD},
\enq
where the ideal part is given by
\beq
\hat{H}_0=\sum_{{\bf i}} 
-\mu{\hat \psi}_{\sigma,{\bf i}}^\dagger{\hat \psi}_{\sigma,{\bf i}}
-\sum_{\sigma,\alpha}\sum_{<{\bf i},{\bf j}>_\alpha} t_{\alpha,\sigma}
{\hat \psi}_{\sigma,{\bf i}}^\dagger{\hat \psi}_{\sigma,{\bf j}}.
\enq
Here 
$\sum_{<{\bf i},{\bf j}>_\alpha}$ indicates the sum over nearest neighbors in the direction $\alpha\in\{x,y,z\}$. Since the Bloch functions diagonalize the single particle Hamiltonian, there
are no interband hopping terms in the Wannier representation considered
here \cite{Georges2007a}. The terms originating from interatomic interactions
are given by
\begin{equation}
\begin{array}{lll}
\hat{H}_{nn} & = & \displaystyle{\sum_{\bf i}\sum_\sigma \frac{U_{\sigma\sigma}}{2}
{\hat n}_{\sigma,\bf i}\left({\hat n}_{\sigma,\bf i}-1\right)} \\ \\
& &
 \displaystyle{+\!\!\,\,\sum_{\bf i}\sum_{\sigma\sigma',\sigma\neq\sigma'}\!\!
U_{\sigma\sigma'}{\hat n}_{\sigma,\bf i}{\hat n}_{\sigma',\bf i}}
\end{array}
\label{eq:Hnn}
\end{equation}
and
\begin{equation}
\begin{array}{lll}
\hat{H}_{FD} & = & \displaystyle{\sum_{\bf i}\sum_{\sigma\sigma',\sigma\neq\sigma'} 
\frac{U_{\sigma\sigma'}}{2}\left({\hat \psi}_{\sigma,{\bf i}}^\dagger {\hat \psi}_{\sigma,{\bf i}}^\dagger {\hat \psi}_{\sigma',{\bf i}} {\hat \psi}_{\sigma',{\bf i}}\right.}\\ \\
& & \displaystyle{+\left.{\hat \psi}_{\sigma',{\bf i}}^\dagger {\hat \psi}_{\sigma',{\bf i}}^\dagger {\hat \psi}_{\sigma,{\bf i}}{\hat \psi}_{\sigma,{\bf i}}
\right)},
\end{array}
\end{equation}
where $\hat{H}_{FD}$ contains terms that describe flavor changing
collisions which transfer atoms between bands. This term has a formal 
similarity with terms responsible
for spin-dynamics in a spinor condensates~\cite{Law1998a,Stamper-Kurn1998b}. 
However, the strength
of these terms is comparable to other interaction terms
as opposed to spinor condensates where it is usually
small, being proportional to the difference between singlet
and triplet scattering lengths (for spin-$1$ spinor condensate).

It should be kept in mind that there are circumstances when
nearest neighbor interactions~\cite{Scarola2005a} or
particle assisted tunneling processes~\cite{Duan2008a} might give 
rise to new physics. These contributions are not
included in the formulation presented here where our focus 
is in the most typical parameter regimes.

The various coupling strengths in the lattice model are
related to $g$ through
\beq
U_{\sigma\sigma'}=g\int d{\bf r} w_{\sigma,{\bf i}}({\bf r})^2w_{\sigma',{\bf i}}({\bf r})^2
\enq
and the tunneling coefficients are given by
\beq
t_{\sigma,\alpha}=-\int d{\bf r} w_{\sigma,{\bf i}}({\bf r})
\left[-\frac{\hbar^2\nabla^2}{2m}+V({\bf r})\right]
w_{\sigma,{\bf i+1}_\alpha}({\bf r}),
\enq
where by ${\bf i+1}_\alpha$ we indicate the neighboring site of ${\bf i}$ in the direction $\alpha$.
When the lattice is symmetric, the tunneling strength on the lowest
band is independent of direction. However, this is not true for the $p$-band where the directional
dependence of the tunneling strength must be kept, as the overlap integrals are very different
depending on whether one is integrating along the node of the Wannier function or orthogonal to it. This indeed has important consequences for the physics in these systems~\cite{Isacsson2005a,Liu2006a,Martikainen2008a}

It should be further noted, that since the parameters of our model are computed using real Wannier functions, we find some, not only quantitative, but also qualitative differences from the commonly used
models build on the harmonic approximation. In particular, many degeneracies appearing in the harmonic approximation are absent when real parameters are used.

\subsection{Validity of $p$-band single-band approximation}
In a harmonic potential, two atoms on the first excited states
have an energy $2\times \hbar\omega(3/2+1)$. This is equal
to the energy of one atom on the ground state and one atom on the 
second excited state. This suggest that collisions between $p$-band atoms
can populate also the lowest $s$-band and the $d$-bands. This would clearly restrict
the validity of the models residing purely on the $p$-bands.

However, a real site in an optical lattice is not exactly
harmonic and this anharmonicity implies that the above processes are normally
off-resonant. The deviation between the real lattice potential
and the harmonic approximation is given by
\beq
\begin{array}{lll}
\Delta V & = & V_L\left[\sin^2 (\pi x/d)+\sin^2 (\pi y/d)+\sin^2 (\pi z/d)\right]
\\ \\
& & \displaystyle{-V_L\pi^2\left[\left(\frac{x}{d}\right)^2+\left(\frac{y}{d}\right)^2+
\left(\frac{z}{d}\right)^2\right].}
\end{array}
\enq
In the first order perturbation theory around the harmonic
approximation, we find that in the limit of deep lattices 
($V_L/E_R\gg 1$, where $E_R$ is the recoil energy) 
the detuning $2E_{1,0,0}-E_{0,0,0}-E_{1,1,0}$, where subsripts denote
quantum numbers $\nu_x$, $\nu_y$, and $\nu_z$ of the harmonic oscillator states,
vanishes. This would be related to a process where two atoms from the $p$-band
scatter into one ground state atom and one atom occupying the 
state $|\nu_x=1,\nu_y=1,\nu_z=0\ra$.
Even though this detuning remains zero at first order in anharmonicity,
the process has a vanishing matrix element and can therefore be ignored.

On the other hand, a process where two atoms from the $p$-band
scatter into one ground state atom and one atom on the state 
$|\nu_x=2,\nu_y=0,\nu_z=0\ra$ (for example) can occur. For this process
the detuning $2E_{1,0,0}-E_{0,0,0}-E_{2,0,0}$ approaches a constant
value of $-2/3\, E_R$ in the limit of deep lattices. Note that 
the oscillator energy $\hbar\omega$ has a $\sqrt{V_L}$ dependence in the same limit,
so even though the detuning approaches a constant for deep lattices, it
becomes small relative to the harmonic oscillator energy scale.
From this we can conclude
that as long as the bandwidths and interactions are very small compared to
recoil energy, we can safely ignore $d$-band atoms and processes which would
scatter atoms from the $p$-band to other bands.

We also note that one way to prevent atoms on the $p$-band to populate the $s$-band was outlined in Ref.~\cite{Liu2006a}. Here, fermionic atoms are occupying the lowest band and due to atom-atom interactions, the $p$-band atoms are blocked from occupying the lowest band.

\section{Gross-Pitaevskii approach}
\label{sec:GPsolutions} 
In the mean-field approach we replace the operators
${\hat\psi_{\alpha,{\bf i}}}$ with complex numbers
$\psi_{\alpha,{\bf i}}$. This approximation amounts to a coherent 
state ansatz in each site. In a Fock representation this is given by
\beq
\begin{array}{lll}
|\psi\ra_{\bf i} & = & \displaystyle{\exp\left({-\frac{|\psi_{{\bf i},x}|^2+
|\psi_{{\bf i},y}|^2+|\psi_{{\bf i},z}|^2}{2}}\right)}\\ \\
& & \displaystyle{\times\sum_{(n_x,n_y,n_z)} 
\frac{\psi_{{\bf i},x}^{n_x}\psi_{{\bf i},y}^{n_y}\psi_{{\bf i},z}^{n_z}}{±\sqrt{n_x!n_y!n_z!}}
|n_x,n_y,n_z\ra_{\bf i}},
\end{array}
\enq
where $\psi_{{\bf i},\alpha}=\la{\hat\psi_{{\bf i},\alpha}}\ra$
is the order parameter for the flavor $\alpha$ at site ${\bf i}=(i_x,i_y,i_z)$.
This mean-field approximation is expected to be reasonably accurate in the superfluid phase when interactions are much weaker than the tunneling strengths. In this same regime the effects due to
the $d$-band atoms, can also be safely ignored as long as the tunneling 
strengths are much smaller than the anharmonicity induced 
detuning discussed earlier.
Using the coherent state ansatz we can 
derive the equations
of motion for the order parameters $\psi_\alpha$ from the
Euler-Lagrange equation
\beq
\frac{\partial L}{\partial\psi_{{\bf i},\alpha}^*}-\frac{d}{dt}\left(\frac{\partial L}{\partial\dot\psi_{{\bf i},\alpha}^*}\right)=0,
\enq
with the Lagrangian given by
\beq
L=\sum_{{\bf i},\alpha}
i\frac{\hbar}{2}\left[\psi_{{\bf i},\alpha}^*\dot\psi_{{\bf i},\alpha}-
\psi_{{\bf i},\alpha}\dot\psi_{{\bf i},\alpha}^*\right]
-H_{MF}.
\enq
Here $H_{MF}$ is the mean-field approximation
for the Hamiltonian in terms of the coherent state amplitudes. 
What we find are the discretized versions
of the Gross-Pitevskii equation for each flavor. These equations
are non-linear and coupled, but can be solved numerically without
too much difficulty. Furthermore, in some special cases analytical results
can even be derived. We choose the lowest band tunneling strength as our unit
of energy and lattice spacing as our unit of length.
Then, for a three-dimensional lattice, the Gross-Pitaevskii
equations for different $p$-band flavors read
\begin{equation}
\begin{array}{lll}
i\hbar\frac{\partial\psi_{{\bf i},x}}{\partial t}&=&
-\sum_\alpha t_{x,\alpha}\left[\psi_{{\bf i+1_\alpha},x}-2\psi_{{\bf i},x}+\psi_{{\bf i-1_\alpha},x}\right]\\ \\
& & +\left[g_{xx}|\psi_{{\bf i},x}|^2+2g_{xy}|\psi_{{\bf i},y}|^2+2g_{xz}|\psi_{{\bf i},z}|^2\right]\psi_{{\bf i},x}\\ \\
& & +\frac{g_{xy}}{2}\psi_{{\bf i},y}^2\psi_{{\bf i},x}^*
+\frac{g_{xz}}{2}\psi_{{\bf i},z}^2\psi_{{\bf i},x}^*,
\end{array}
\end{equation}
\begin{equation}
\begin{array}{lll}
i\hbar\frac{\partial\psi_{{\bf i},y}}{\partial t}&=&
-\sum_\alpha -_{y,\alpha}\left[\psi_{{\bf i+1_\alpha},y}-2\psi_{{\bf i},y}+\psi_{{\bf i-1_\alpha},y}\right]\\ \\
& & +\left[g_{yy}|\psi_{{\bf i},y}|^2+2g_{xy}|\psi_{{\bf i},x}|^2+2g_{yz}|\psi_{{\bf i},z}|^2\right]\psi_{{\bf i},y}\\ \\
& &+\frac{g_{xy}}{2}\psi_{{\bf i},x}^2\psi_{{\bf i},y}^*
+\frac{g_{yz}}{2}\psi_{{\bf i},z}^2\psi_{{\bf i},y}^*,
\end{array}
\end{equation}
and
\begin{equation}
\begin{array}{lll}
i\hbar\frac{\partial\psi_{{\bf i},z}}{\partial t}&=&
-\sum_\alpha t_{z,\alpha}\left[\psi_{{\bf i+1_\alpha},z}-2\psi_{{\bf i},z}+\psi_{{\bf i-1_\alpha},z}\right]\\ \\
& & +\left[g_{zz}|\psi_{{\bf i},z}|^2+2g_{xz}|\psi_{{\bf i},x}|^2+2g_{yz}|\psi_{{\bf i},y}|^2\right]\psi_{{\bf i},z}\\ \\
& &+\frac{g_{xz}}{2}\psi_{{\bf i},x}^2\psi_{{\bf i},z}^*
+\frac{g_{yz}}{2}\psi_{{\bf i},y}^2\psi_{{\bf i},z}^*.
\end{array}
\end{equation}
In these equations the first term on the right hand side
is due to the kinetic energy in the lattice, the second term
originates from the density-density interactions, while the
last terms are due to the flavor changing collisions.
The generalization for the two-dimensional system with only
two flavors is straight forward.

\subsection{2-dimensional lattice}
In a two-dimensional system we have two degenerate $p$-bands.
On a mean-field level it is easy to investigate
the lowest energy wavefunctions in the broken symmetry phase.
When the lattice is very deep, the energy minimization can be done
in each site separately by ignoring the tunneling term entirely. 
In this way we find that the lowest energy state in each site
is given by $\psi_x=e^{i\phi}/\sqrt{2}$ and $\psi_y=e^{i\phi\pm \pi/2}/\sqrt{2}$.
This corresponds to an onsite wavefunction
\beq
\la {\hat\psi}({\bf x})\ra=w_x({\bf x})\psi_x+w_y({\bf x})\psi_y.
\enq
Since the Wannier functions are related to each other
and can be expressed as $w_x({\bf x})=f(x)w_0({\bf x})$
and $w_y({\bf x})=f(y)w_0({\bf x})$, this implies
\beq
\la {\hat\psi}({\bf x})\ra=\frac{e^{i\phi}}{\sqrt{2}}w_0({\bf x})
\left(f(x)\pm i f(y)\right).
\enq
For deep lattices the Wannier functions approach the harmonic oscillator
states and $f(x)\sim x$. We can then clearly see that the mean-field
state corresponds to a vortex or anti-vortex state with an angular momentum
$\pm 1$ along the $z$-axis.

Within this approximation, any configuration of either vortex or anti-vortex states at each site are degenerate. However, when the tunneling term is non-zero, phases of the order parameters
in different sites must be correlated properly if the energy is to be minimized.
When tunneling strengths are positive, the lowest energy
condensed state has the same phase at each site. However,
on the $p$-band the tunneling strength for a flavor is negative
in the direction of the node in its localized Wannier function.
In this case, the lowest energy state has a $\pi$ phase-difference
between neighboring sites. For a two-dimensional system
it is possible to find the mean-field state which minimizes
the onsite problem as well as the tunneling problem simultaneously
and this state amounts to a checkerboard (or anti-ferromagnetic)
ordering of vortices and anti-vortices. 

This is easy to see, since if at some site we have a vortex
state $\sim (x+iy)$ and we aim to minimize the kinetic energy
along $y$-direction, then the neighboring site should have
a same phase for the $x$-flavor while having a $\pi$-phaseshift
for the $y$-flavour. This implies an anti-vortex state $\sim (x-iy)$.
If we then try to minimize the kinetic energy along $x$-direction,
we see that the $x$-flavor should experience a $\pi$-phaseshift,
while for the $y$-flavour the phaseshift should vanish.
This implies an anti-vortex state $\sim e^{i\pi} (x-iy)$ with an
additional overall phaseshift of $\pi$.

\subsection{3-dimensional lattice}
In a three-dimensional lattice we have three degenerate bands, which opens up for novel phenomena not present in the two-dimensional case. Mimimizing the onsite problem we find that the lowest energy configuration becomes
\beq\label{120deg}
\la\Psi\ra=\left(\begin{array}{c}\la\psi_x\ra\\
\la\psi_y\ra\\
\la\psi_z\ra
\end{array}\right)=\sqrt{\frac{n_T}{3}}e^{i\phi}\left(\begin{array}{c} 
1\\
\exp(2\pi i/3)\\
\exp(4\pi i/3)
\end{array}\right),
\enq
where $n_T$ is the total onsite atom number and $\phi$ is a random phase. 
The onsite wavefunction with equal number of atoms in each flavor
has an unit angular momentum per atom which points
not along the main axes, but diagonally ${\bf L}\propto (\pm 1,\pm 1,\pm1)$. Again minimization of the kinetic energy necessitates a special ordering of angular momentum in each site. 
In the three-dimensional lattice the nearest neighbor angular momenta (in the direction $\alpha$) are related by a relation ${\bf L}({\bf i}+{\bf e}_\alpha)=\hat{R}_\alpha(\pi){\bf L}({\bf i})$, where
$\hat{R}_\alpha(\pi)$ is a rotation of $\pi$ around the axis $\alpha$.

The above results depend crucially on the magnitude of the inter-flavor coupling strengths $g_{xy}=g_{xz}=g_{yz}$ relative to the magnitudes of the $g_{\alpha\alpha}$ terms. In particular, it only holds when $g_{xy}< g_{xx}/3$. If one approximates the Wannier functions with the harmonic oscillator states one  finds that $g_{xy}=g_{xx}/3$, but when real Wannier functions of an ideal
Bose gas are used, $g_{xy}<g_{xx}/3$ for fairly deep lattices and the above result holds. 
That said, the result may be different for a shallow lattice. Furthermore, it is also unclear 
what is the effect of the interaction-induced dressing of the Wannier functions~\cite{Li2006a} on the magnitudes of the effective $p$-band coupling strengths.
If it turns out that under some circumstances inter-flavor
coupling is larger and $g_{xy}>g_{xx}/3$, then the lowest energy configuration
breaks the permutational symmetry and is given by the vortex states
\beq\label{90deg}
\la\Psi\ra=\sqrt{\frac{n_T}{2}}e^{i\phi}\left(\begin{array}{c} 
1\\
\exp(\pm \pi i/2)\\
0
\end{array}\right),
\enq
where the angular momentum points along the $z$-axis. The vortex-anti-vortex states
with angular momentum along other axes are degenerate with the one shown here explicitly. It is seen that the state~(\ref{120deg}) has a mutual $120^\circ$ phase difference between the flavors, reminiscent of three interacting spin 1/2-particles placed on the corners of a triangle. The state (\ref{90deg}), on the other hand, shows a mutual $90^\circ$ phase pattern. In particular, the interaction terms proportional to $g_{\alpha\beta}$, with $\alpha=\beta$, favors a $90^\circ$ pattern, while those with $\alpha\neq\beta$ favor a $120^\circ$ configuration. 

In order to achieve a better understanding how the particular limiting case $g_{xy}=g_{xx}/3$ comes about, let us again write the Hamiltonian as $H=H_{nn}+H_{FD}$ where in the mean-field approximation we have
\begin{equation}\label{en1}
\begin{array}{lll}
H_{nn} & = & g_{xx}\left[n_x^2+n_y^2+n_z^2\right]\\ \\
& & +4g_{xy}(n_xn_y+n_xn_z+n_yn_z),\\ \\
H_{FD} & = & 2g_{xy}\left[\cos(\Delta_{xy})n_xn_y+\cos(\Delta_{xz})n_xn_z\right.\\ \\
& & \left.+\cos((\Delta_{xz}-\Delta_{xy}))n_yn_z\right].
\end{array}
\end{equation}
Here, $n_\alpha=|\psi_\alpha|^2$, and $\Delta_{\alpha\beta}=\phi_\alpha-\phi_\beta$, with $\phi_\alpha$ being the phase of $\langle\psi_\alpha\rangle$. The energy functional can be written in the form
\begin{equation}
E[\langle\psi_x\rangle,\langle\psi_y\rangle,\langle\psi_z\rangle]=\mathbf{n}^T\mathbf{M}\mathbf{n},
\end{equation}
where $\mathbf{n}=(n_x,n_y,n_z)$ and 
\begin{widetext}
\begin{equation}
\mathbf{M}=\left[\begin{array}{ccc}
g_{xx} & 2g_{xy}\left(2+\cos(\Delta_{xy})\right) & 2g_{xy}\left(2+\cos(\Delta_{xz})\right)\\
2g_{xy}\left(2+\cos(\Delta_{xy})\right) & g_{xx} & 2g_{xy}\left(2+\cos((\Delta_{xz}-\Delta_{xy}))\right)\\
2g_{xy}\left(2+\cos(\Delta_{xz})\right) & 2g_{xy}\left(2+\cos((\Delta_{xz}-\Delta_{xy}))\right) & g_{xx}
\end{array}\right].
\end{equation}
\end{widetext}
Thus, we have rewritten the single site problem in a quadratic form for the $n_\alpha$ variables.
For the general case, the eigenvalues are not analytically solvable. However, assuming $g_{xy}<g_{xx}/3$ we may use the fact that we know that the energy is minimized for $\Delta_{xy}=2\pi/3$ and $\Delta_{xz}=4\pi/3$ and we then obtain
\begin{equation}\label{pos}
\begin{array}{l}
\lambda_1=g_{xx}-3g_{xy},\\
\lambda_2=g_{xx}-3g_{xy},\\
\lambda_3=g_{xx}+6g_{xy}.
\end{array}
\end{equation}
Since $g_{xx}>3g_{xy}$, the matrix $\mathbf{M}$ is positive definite. However, putting $g_{xx}<3g_{xy}$ into Eq.~(\ref{pos}) results in a non positive definite matrix and we can conclude that the $120^\circ$ phase symmetry is broken in such a case. The possibility of the broken permutational symmetry was also
noted in Ref.~\cite{Isacsson2005a}.

We should point out that all the results rely on having an isotropic lattice configuration. Any deviation from the symmetric lattice will break this degeneracy and give a preferred direction for the axis of angular momentum. When kinetic energy is included the ordering of vortex anti-vortex states between sites is the same as in the two-dimensional system.

\section{Quantum states}
\label{sec:Gutzwiller2D}

The Gross-Pitaevskii mean-field approximation usually provides a sufficient description 
when one considers the superfluid phase with a large number of atoms per site. 
However, the local coherent state ansatz is not necessarily all 
that good when the average onsite occupation number is small. 
Furthermore, the mean-field description fails completely when the system is in a
Mott insulator phase {\it i.e.} when the onsite atom distribution is greatly sub-Poissonian. Therefore, to accurately describe the system properties in this regime a more precise many-body wave function is needed. This will also provide insight into which parameter regimes where the mean-field picture  
is a good approximation. 

We assume that the state vector has the generalized form of the Gutzwiller 
approximation~\cite{Buonsante2008a}. This is a product of 
on-site quantum states expanded in terms of the Fock states $|{\bf n}\ra$
of the multiple flavor system
\beq
\label{eq:Gutzwiller}
|\psi\ra=\prod_{\bf i} \sum_{\bf n }f_{{\bf n}}^{({\bf i})} |{\bf n}\ra_{\bf i}
\enq
where the index ${\bf i}$ runs over all lattice sites. The expansion 
coefficient $f_{{\bf n}}^{({\bf i})}$ is the Gutzwiller amplitude 
of the particular on-site Fock state. For our purposes, in
the $p$-band the relevant subspace is covered by the Fock states of the form 
$|{\bf n}\ra=|n_x,n_y,n_z\ra$, where for example, $n_x$ is the occupation
number of the $p_x$-flavor. 

Within the Gutzwiller approximation the energy of the system becomes 
a functional of the amplitudes $f_{{\bf n}}^{({\bf i})}$. By utilizing a conjugate gradient method we minimize this functional giving the system ground state at zero temperature, $T=0$. Several aspects regarding the minimization were discussed in Ref. \cite{Larson2009a}. Here we only 
mention that the sum over ${\bf n}$ must be cut off and in our numerical
scheme we include all the states with $\sum n_{\sigma} \le 8$ in 2D and 
$\sum n_{\sigma} \le 6$ in 3D. Throughout this section we will choose the lattice amplitude to be fixed and instead assume that the ratio between tunneling and onsite interaction can be controlled via Feshbach resonances. The Wannier functions are calculated for a relatively deep lattice, $V_L=15E_R$

\subsection{2-dimensional lattice}
\label{sec:2d-gutzw}

For the two-dimensional lattice, the onsite Fock states then consist of $p_x$ and $p_y$ terms only. To get insight about possible correlations between neighboring lattice sites our effective computational subspace contains four lattice sites with two sites in both spatial directions. The computational $4t/U_{00}$-$\mu/U_{00}$ 
parameter region is chosen to be such that the total number of 
atoms per site is relatively small. To investigate the effects of quantum fluctuations, this is the region of most experimental interest and it is also favorable numerically with reasonable 
cut-offs.

\begin{figure}
\includegraphics[width=8cm]{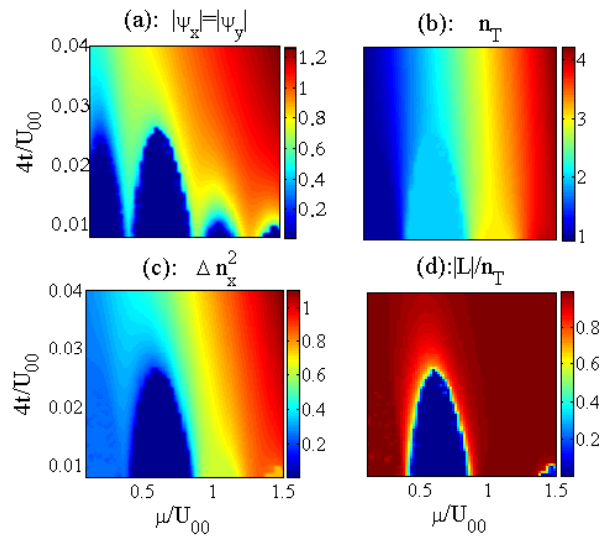}
\caption{(Color online) Properties of the two-dimensional two-flavor
Bose-Hubbard model as a function of the chemical potential and the inverse
interaction strength $4t/U_{00}$ where the factor 4 derives from the number
of nearest neighbors. For concreteness the parameters were computed for
a lattice depth of $V_L=15E_R$. The various plots show: order parameters (a), total atom number (b), atom number fluctuations (c), and angular momentum per particle (d).}
 \label{fig:properties1}
\end{figure}

Some resulting properties can be seen in Fig. \ref{fig:properties1}. 
The absolute value of the two condensate order parameters $|\langle\psi_x\rangle|,\,|\langle\psi_y\rangle|$ are plotted in Fig. \ref{fig:properties1} (a). As in the standard $s$-band Bose-Hubbard model, the phase-space consists of Mott insulating lobes and superfluid regions. This is further evidenced in (b) where the total atom number $n_T$ is shown; within the Mott lobes $n_T$ attains an integer value. Due to symmetry reasons, it is not surprising that the absolute values of the two flavor order parameters are identical. However, in the SF phase, there is a phase difference of $\pm\pi/2$ between the two flavors e.g., $\langle\psi_x\rangle$ is real while $\langle\psi_y\rangle$ is imaginary. This suggests that the on-site ground state is a vortex; a result in agreement with our Gross-Pitaevskii calculations. Indeed, a plot of the scaled angular momentum $|L|/n_T\equiv\hat{L}_{z,{\bf i}}=-i\left(\hat{\psi}_{x,{\bf i}}^\dagger\hat{\psi}_{y,{\bf i}}-\hat{\psi}_{y,{\bf i}}^\dagger\hat{\psi}_{x,{\bf i}}\right)/n_T$ given in Fig.~\ref{fig:properties1} (d) verifies that for strong tunnelings and large onsite atom numbers the angular momentum is quantized. The existence of the vortex solution is also supported by the work of Watanabe and Pethick~\cite{Watanabe2007}. Namely, in a single harmonic trap within the the mean field approximation the energy functional is of the form 
\beq
E_{MFHO}=
-\frac{\gamma}{4}\left[1-\left(\Delta n\right)\right]
\sin^2\phi
\label{eq:mfho}
\enq
where $\gamma$ is the effective coupling constant, $\Delta n = n_x-n_y$ 
population difference between the two flavors, and $\phi$ is their relative phase. Eq. (\ref{eq:mfho}) is clearly minimized when $\Delta n=0$ and $\phi=\pm\pi/2$. Physically this means that the repulsive interaction favors a 
vortex solution above a non-vortex one. 

As discussed in the previous section, in the mean-field limit the 
vortices on the lattice tend to order themselves in a 
form of a checkerboard pattern with neighboring vortices and anti-vortices. 
According to our Gutzwiller results this is true also more 
generally in the superfluid phase. In fact, our ground state of 
anti-ferromagnetic like vortex ordering is similar to the 
staggered-vortex superfluid state discussed in 
Ref.~\cite{Lim2008a} for a square optical lattice in an effective 
staggered magnetic field. However, in the $p$-band such a state 
appears even in the absence of effective magnetic fields.

The physics appearing for the multi-flavor Mott insulating states is possibly even more interesting. For example, as seen in Fig.~\ref{fig:properties1} (c), onsite number fluctuations $\Delta n_x^2$ (or equivalently $\Delta n_y^2$) for the individual flavors are not necessarily zero. For the Gutzwiller ansatz wave function~(\ref{eq:Gutzwiller}), no correlation between sites is allowed. As an outcome, for odd total number of atoms $n_T$ there is a set of degenerate Mott states, {\it e.g.} with $n_T=1$ all onsite interaction terms vanish and the state $|n_x=1,n_y=0\rangle$ is degenerate with $|n_x=0,n_y=1\rangle$ or any linear combination of these. However, tunneling between sites will normally break these degeneracies. The Gutzwiller approach is not able to capture such effects and therefore the kinetic energy term
\begin{equation}
\hat{T}=\sum_{\sigma,\alpha}\sum_{<{\bf i},{\bf j}>_\alpha}t_{\alpha,\sigma}\hat{\psi}_{\sigma,{\bf i}}^\dagger\hat{\psi}_{\sigma,{\bf j}}
\end{equation}
is taken into account within second order perturbation theory. We focus on the lowest Mott, $n_T=1$, and it turns out that the degeneracy is indeed lifted and the ground state shows a anti-ferromagnetic vortex structure. We note that the 
favorability of a vortex state in the $n_T=1$ Mott state relies to the
non zero value of the transverse tunneling rate. If this tunneling 
is compeletely neglected the energy is 
minimized by a ferromagnetic state \cite{Isacsson2005a}. In the second 
lowest Mott, $n_T=2$, the picture is simpler because the interactions break 
the degeneracy and no perturbation theory is needed. 
The state $|n_x=1,n_y=1\rangle$ is favored over $|n_x=2,n_y=0\rangle$ and
$|n_x=0,n_y=2\rangle$ due to the vanishing of the self terms proportional to  
${\hat n}_{\sigma,\bf i}\left({\hat n}_{\sigma,\bf i}-1\right)$ in Eq. 
(\ref{eq:Hnn}).

\begin{figure}
\includegraphics[width=8cm]{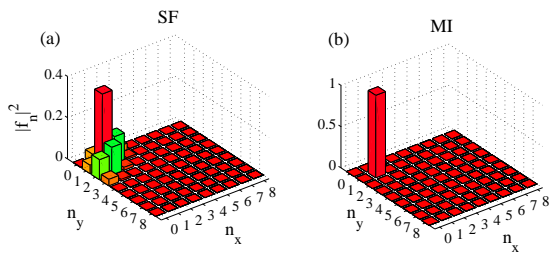} 
\caption{(Color online) Absolute values of the Gutzwiller amplitudes at a 
single lattice cite. The left figure (a), shows the atomic distribution
for a superfluid ground state with $4t/U_{00}=0.04$ 
and $\mu/U_{00}=0.7$. On the right figure (b), a Mott insulator state is plotted
for $4t/U_{00}=0.01$ and $\mu/U_{00}=0.7$. Expectedly, in this insulator phase 
only the Fock state $|n_x,n_y \rangle=|1,1 \rangle$ is populated within the Gutzwiller approach.}
\label{fig:absamplitudes}
\end{figure}

We further illustrate our results by plotting the absolute values of the Gutzwiller amplitudes $f_{\bf n}^{({\bf i})}$ in Fig. \ref{fig:absamplitudes} as bar graphs.  Hence, these are the probabilities of the onsite state to be at a given 
Fock state $|n_x,n_y \rangle$. In Fig. \ref{fig:absamplitudes} (a) 
the single-site amplitudes of a superfluid state are 
given when $4t/U_{00}=0.04$ and $\mu/U_{00}=0.7$. This state is 
clearly a superposition of many Fock states whereas the 
Mott state of Fig. \ref{fig:absamplitudes} (b) contains only one state. 
This MI state is the minimum energy configuration for   
 $4t/U_{00}=0.01$ and $\mu/U_{00}=0.7$. Due to the small number of atoms, the superfluid atomic distribution depicted in Fig.~\ref{fig:absamplitudes} (a) is still sub-Poissonian. It is also evident from the figures that the states with extensive populations are negligible justifying our numerical cut-off at $n_\sigma\leq8$.

\subsection{3-dimensional lattices}
\label{sec:Gutzwiller3D}
In a symmetric three dimensional lattice the $p$-band is described in terms of $3$-flavors.
In the Mott insulator with only one atom per site, for the same reason as for the two-dimensional case, the ground state is strongly degenerate within the Gutzwiller ansatz. As argued above, such states are not true eigenstates of our Bose-Hubbard Hamiltonian, and again for relatively deep lattices the breaking of this degeneracy, and hence the permutational symmetry breaking, is well described within second order perturbation theory. Using real Wannier functions to compute the model parameters we find that, in a theory which takes the kinetic energy into account perturbatively, the ferromagnetic state where only one 
flavor is occupied has a lower energy than either anti-ferromagnetic states with checkerboard ordering or striped phases.


\begin{figure}
\includegraphics[width=8cm]{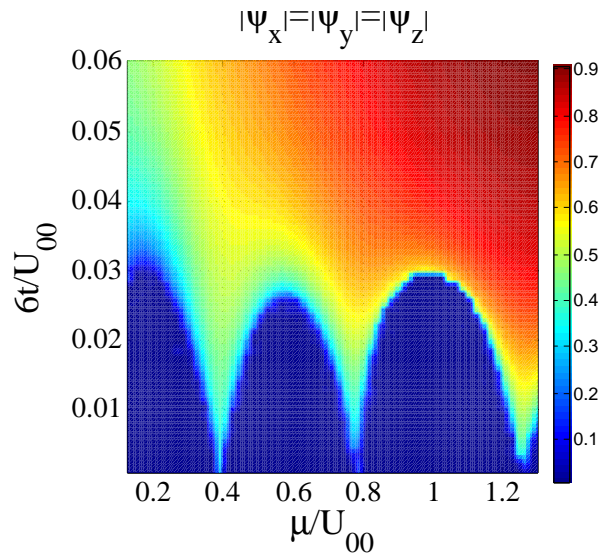} 
\caption{(Color online) The condensate order parameters for the three-dimensional lattice.
}
\label{Fig3}
\end{figure}

With only two atoms per site the condensate order parameters
naturally vanish when entering the Mott insulating regime, but the local
angular momentum 
$\la \hat{L}\ra=\frac{1}{\sqrt{3}}\left(\pm 1,\pm 1,\pm 1\right)$
is non-zero and
$\la L^2\ra$ is equal to $6$. Angular momentum per particle
$\sqrt{\sum_\alpha \la\hat{L}_\alpha\ra^2}/n_T$ is $1/2$ in this state
and is in a marked contrast to the superfluid regime, where
the onsite angular momentum per particle is equal to one.
In a superfluid phase, the half-quantum vortex 
can occur in multi-component systems and 
can be pictured 
as a vortex in one of the component with the vortex free component
filling the vortex core~\cite{Leonhardt2000a,Martikainen2002b}.
However, being non-zero the expectation value of the angular momentum
is in qualitative agreement with the Gross-Pitaevskii solution
even in the Mott lobe.
More explicitly, the minimum energy state
in each site is maximally entangled angular momentum eigenstate 
given by
\beq\label{symstate}
|\psi\ra\frac{1}{\sqrt{3}}\left[e^{i\phi_1}|110\ra
+e^{i\phi_2}|101\ra+e^{i\phi_3}|011\ra\right],
\enq
where the amplitudes have $2\pi/3$ phase-differences. 

For three atoms per site, the lowest energy Mott insulator 
state has the onsite wavefunction $\psi=|111\ra$, which was also found for the corresponding state in the two-dimensional lattice. Importantly it should be
noted, that commonly used harmonic approximation
for the Wannier states 
predicts the properties of (for example) this insulating phase incorrectly.
If harmonic oscillator states are used to approximate
Wannier wavefunctions, the insulating state
with $3$ atoms per site is degenerate with more complicated 
superposition states, but these degeneracies are removed once real
Wannier states are used to evaluate the parameters of the theory.
As we have demonstrated, tunneling between sites will remove the onsite degeneracies among the Mott insulating states, a fact that was already pointed out by Isacsson {\it et al.}~\cite{Isacsson2005a}. However, many of the degeneracies appearing in their work are actually artifacts of utilizing a harmonic approximation. Furthermore, we found that the $p$-band Mott lobes follow roughly the structure for the Mott lobes on the lowest band, as depicted in Figs.~\ref{fig:properties1} and~\ref{Fig3}. This is in contrast with the results of Ref.~\cite{Isacsson2005a} where the Mott lobes extends over larger parameter regimes, and they moreover show an anomalous behavior with large variations in the sizes of neighboring Mott lobes. This discrepancy seems to originate from a factor of $2$ missing for the cross terms proportional to $n_xn_y$, $n_yn_z$, and $n_yn_z$ in their work.

\begin{figure}
\includegraphics[width=0.90\columnwidth]{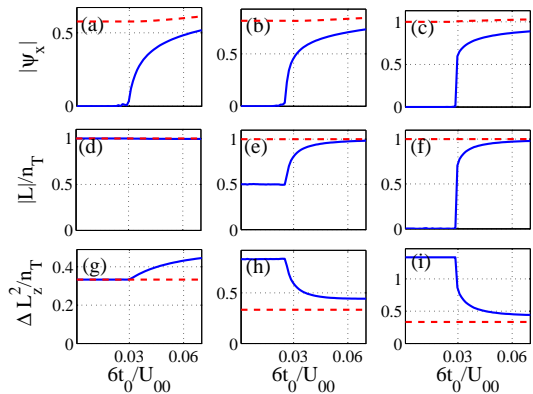} 
\caption{(Color online) Comparison between the Gutzwiller approach (solid blue lines) and
the Gross-Pitaevskii theory (dashed red lines) in a three-dimensional system. 
The parameters were computed for a lattice of depth $15\,{\rm E_R}$
in all directions. We fix $\mu/U_{00}$ in
the Bose-Hubbard model and changed $6t_{0}/U_{00}$. 
We show comparisons for the condensate order parameter $\la\psi_x\ra$,
onsite angular momenta per particle $|L|/n_T$, as well as
for the fluctuations $\Delta L_x^2/n_T$.
In (a), (d), and (g) the strong coupling region was in a Mott state
with $1$ atom per site,
in (b), (e), and (h) the strong coupling region was in a Mott state
with $2$ atoms per site, and in 
(c), (f), and (i) the strong coupling region was in a Mott state
with $3$ atoms per site. Note that we choose a specific Mott insulating
state with $n_T=1$ so that it had an angular momentum of $1$ per atom.
Since this region is in our approximation
strongly degenerate, many other
choices would have been equally justified.
}
\label{Fig:GutzvsGP_3D}
\end{figure}

In Fig.~\ref{Fig:GutzvsGP_3D} we compare the Gutzwiller approach and the Gross-Pitaevskii approach by showing one component of the condensate order parameters, angular momenta per particle, as well
as the fluctuations of the $z$-component of the angular momentum as a function of $6t_{0}/U_{00}$.
In this figure we fixed $\mu/U_{00}$ in the Bose-Hubbard model phase diagram and changed $6t_{0}/U_{00}$ 
and for each point computed the corresponding solution of the Gross-Pitaevskii equations with the same density. The fixed values of $\mu/U_{00}$ were chosen in such a way that
the starting point was in the center of the Mott insulating phase with either
$1$, $2$, or $3$ atoms per site.

With the exception of fluctuations of the single particle per site angular momentum, we can see that outside the Mott insulating regions the Gross-Pitaevskii
theory can quickly predict the value of the condensate order parameters
quite accurately. Angular momenta are in a sense sometimes
even better predicted by the Gross-Pitaevskii theory, since
in the Mott phase with $2$ atoms per site angular momentum is non-zero
and behaves qualitatively in the same way in the two different
approaches. However, with $3$ atoms per site the angular momentum vanishes in the Gutzwiller approach, but is non-zero in the Gross-Pitaevskii
approach. Also the fluctuations of angular momentum agree well in the SF regime. These results give us a benchmark for the reliable use of the Gross-Pitaevskii formalism for the description of the
excited band bosons, and we especially find that the mean-field treatment is surprisingly accurate even relatively close to the Mott boundaries were quantum fluctuations are known to become significant.

Earlier we pointed out the possibility of the broken permutational symmetry when $g_{xx}>3g_{xy}$.
In this case the order parameters can be unequal. Interestingly, we find that this broken symmetry is also reflected in the Mott insulating state, where
the exact ground state (with two atoms per site in this example) 
carrying angular momentum changes into  a superposition
\beq\label{asymstate}
|\psi\ra\frac{1}{\sqrt{3}}\left[\sqrt{p_x}e^{i\phi_1}|200\ra
+\sqrt{p_y}e^{i\phi_2}|020\ra+\sqrt{p_z}e^{i\phi_3}|002\ra\right]
\enq
with possibly unequal number of atoms in different flavors, in contrast to the symmetric state (\ref{symstate}). In particular, for the state (\ref{asymstate}) the angular momentum vanishes.
As one  moves to the superfluid phase from the Mott phase, the permutational symmetry breaking can
manifest itself by a single non-vanishing order parameter $\la\psi_\alpha\ra$ followed by a transition into a state with two non-vanishing (and equal) order parameters~\cite{Isacsson2005a}.

\section{Conclusions}
\label{sec:conclusions}
In this paper we have explored the properties of bosonic atoms
on the first excited band of an optical lattice. By computing 
the phase diagrams for two- and three-dimensional systems, we 
found Mott-insulating and superfluid phases with more subtle
quantum properties than those appearing in the lowest band Hubbard model.
Furthermore, we compared the Gutzwiller theory to the Gross-Pitaevskii
approach and established the parameter regimes where the latter
description provides a good approximation to the physical system.

Here we found that bosons on the $p$-band can form
a staggered-vortex superfluid composed of anti-ferromagnetically ordered
vortices and anti-vortices. Rotation breaks the degeneracy of the
vortex and anti-vortex state and it would be interesting to
explore how rotation favoring vortex lattice formation competes
with the physics of staggered-vortex superfluids. Also, in fairly shallow lattices where effects due 
to interactions can be pronounced, dispersions can develop swallowtails
in the vicinity of the Brillouin zone edge and period doubled states can appear~\cite{Machholm2003a,Machholm2004a}. In the previous analysis which assumed an one-dimensional systems, the swallowtails were found to be related to 
the existence of solutions corresponding to a train of solitons. It would be of interest to explore the similar situation in higher dimensions, where stability properties are often very different.

Experiments are typically done in optical lattices with an additional trapping potential acting at the background. Here we studied only the homogeneous solutions and this assumption is valid locally when the background trapping potential varies slowly compared to the lattice spacing. Our results can be applied in a trap using the local density approximation or by adding a site dependent energy offset to the Hamiltonian. However, the size of computations using the multi-flavor Gutzwiller ansatz grow quickly as a function of system size which at this stage limits us to fairly small systems. Inhomogeneous density distribution is easier to take into account within the mean-field approximation.

\bibliographystyle{apsrev}
\bibliography{./bibli}

\begin{thebibliography}{30}
\expandafter\ifx\csname natexlab\endcsname\relax\def\natexlab#1{#1}\fi
\expandafter\ifx\csname bibnamefont\endcsname\relax
  \def\bibnamefont#1{#1}\fi
\expandafter\ifx\csname bibfnamefont\endcsname\relax
  \def\bibfnamefont#1{#1}\fi
\expandafter\ifx\csname citenamefont\endcsname\relax
  \def\citenamefont#1{#1}\fi
\expandafter\ifx\csname url\endcsname\relax
  \def\url#1{\texttt{#1}}\fi
\expandafter\ifx\csname urlprefix\endcsname\relax\def\urlprefix{URL }\fi
\providecommand{\bibinfo}[2]{#2}
\providecommand{\eprint}[2][]{\url{#2}}

\bibitem[{\citenamefont{Bloch et~al.}(2008)\citenamefont{Bloch, Dalibard, and
  Zwerger}}]{Bloch2008a}
\bibinfo{author}{\bibfnamefont{I.}~\bibnamefont{Bloch}},
  \bibinfo{author}{\bibfnamefont{J.}~\bibnamefont{Dalibard}}, \bibnamefont{and}
  \bibinfo{author}{\bibfnamefont{W.}~\bibnamefont{Zwerger}},
  \bibinfo{journal}{Rev. Mod. Phys.} \textbf{\bibinfo{volume}{80}},
  \bibinfo{pages}{885} (\bibinfo{year}{2008}).

\bibitem[{\citenamefont{Lewenstein et~al.}(2007)\citenamefont{Lewenstein,
  Sanpera, Ahufinger, Damski, Sen, and Sen}}]{Lewenstein2007a}
\bibinfo{author}{\bibfnamefont{M.}~\bibnamefont{Lewenstein}},
  \bibinfo{author}{\bibfnamefont{A.}~\bibnamefont{Sanpera}},
  \bibinfo{author}{\bibfnamefont{V.}~\bibnamefont{Ahufinger}},
  \bibinfo{author}{\bibfnamefont{B.}~\bibnamefont{Damski}},
  \bibinfo{author}{\bibfnamefont{A.}~\bibnamefont{Sen}}, \bibnamefont{and}
  \bibinfo{author}{\bibfnamefont{U.}~\bibnamefont{Sen}}, \bibinfo{journal}{Adv.
  Phys.} \textbf{\bibinfo{volume}{56}}, \bibinfo{pages}{243}
  (\bibinfo{year}{2007}).

\bibitem[{\citenamefont{Jaksch et~al.}(1998)\citenamefont{Jaksch, Bruder,
  Cirac, Gardiner, and Zoller}}]{Jaksch1998a}
\bibinfo{author}{\bibfnamefont{D.}~\bibnamefont{Jaksch}},
  \bibinfo{author}{\bibfnamefont{C.}~\bibnamefont{Bruder}},
  \bibinfo{author}{\bibfnamefont{J.}~\bibnamefont{Cirac}},
  \bibinfo{author}{\bibfnamefont{C.~W.} \bibnamefont{Gardiner}},
  \bibnamefont{and} \bibinfo{author}{\bibfnamefont{P.}~\bibnamefont{Zoller}},
  \bibinfo{journal}{Phys. Rev. Lett.} \textbf{\bibinfo{volume}{81}},
  \bibinfo{pages}{3108} (\bibinfo{year}{1998}).

\bibitem[{\citenamefont{Greiner et~al.}(2002)\citenamefont{Greiner, Mandel,
  Esslinger, {H\"{a}nsch}, and Bloch}}]{Greiner2002a}
\bibinfo{author}{\bibfnamefont{M.}~\bibnamefont{Greiner}},
  \bibinfo{author}{\bibfnamefont{O.}~\bibnamefont{Mandel}},
  \bibinfo{author}{\bibfnamefont{T.}~\bibnamefont{Esslinger}},
  \bibinfo{author}{\bibfnamefont{T.~W.} \bibnamefont{{H\"{a}nsch}}},
  \bibnamefont{and} \bibinfo{author}{\bibfnamefont{I.}~\bibnamefont{Bloch}},
  \bibinfo{journal}{Nature} \textbf{\bibinfo{volume}{415}}, \bibinfo{pages}{39}
  (\bibinfo{year}{2002}).

\bibitem[{\citenamefont{K{\"{o}}hl et~al.}(2006)\citenamefont{K{\"{o}}hl,
  G{\"{u}}nter, St{\"{o}}ferle, Moritz, and Esslinger}}]{Kohl2006a}
\bibinfo{author}{\bibfnamefont{M.}~\bibnamefont{K{\"{o}}hl}},
  \bibinfo{author}{\bibfnamefont{K.}~\bibnamefont{G{\"{u}}nter}},
  \bibinfo{author}{\bibfnamefont{T.}~\bibnamefont{St{\"{o}}ferle}},
  \bibinfo{author}{\bibfnamefont{H.}~\bibnamefont{Moritz}}, \bibnamefont{and}
  \bibinfo{author}{\bibfnamefont{T.}~\bibnamefont{Esslinger}},
  \bibinfo{journal}{J. Phys. B: At. Mol. Opt. Phys.}
  \textbf{\bibinfo{volume}{39}}, \bibinfo{pages}{S47} (\bibinfo{year}{2006}).

\bibitem[{\citenamefont{Alon et~al.}(2005)\citenamefont{Alon, Streltsov, and
  Cederbaum}}]{Alon2005a}
\bibinfo{author}{\bibfnamefont{O.~E.} \bibnamefont{Alon}},
  \bibinfo{author}{\bibfnamefont{A.~I.} \bibnamefont{Streltsov}},
  \bibnamefont{and} \bibinfo{author}{\bibfnamefont{L.~S.}
  \bibnamefont{Cederbaum}}, \bibinfo{journal}{Phys. Rev. Lett.}
  \textbf{\bibinfo{volume}{95}}, \bibinfo{pages}{030405}
  (\bibinfo{year}{2005}).

\bibitem[{\citenamefont{Browaeys et~al.}(2005)\citenamefont{Browaeys, Haffner,
  McKenzie, Rolston, Helmerson, and Phillips}}]{Browaeys2005a}
\bibinfo{author}{\bibfnamefont{A.}~\bibnamefont{Browaeys}},
  \bibinfo{author}{\bibfnamefont{H.}~\bibnamefont{Haffner}},
  \bibinfo{author}{\bibfnamefont{C.}~\bibnamefont{McKenzie}},
  \bibinfo{author}{\bibfnamefont{S.~L.} \bibnamefont{Rolston}},
  \bibinfo{author}{\bibfnamefont{K.}~\bibnamefont{Helmerson}},
  \bibnamefont{and} \bibinfo{author}{\bibfnamefont{W.~D.}
  \bibnamefont{Phillips}}, \bibinfo{journal}{Phys. Rev. A}
  \textbf{\bibinfo{volume}{72}}, \bibinfo{pages}{053605}
  (\bibinfo{year}{2005}).

\bibitem[{\citenamefont{M{\"{u}}ller et~al.}(2007)\citenamefont{M{\"{u}}ller,
  F{\"{o}}lling, Widera, and Bloch}}]{Mueller2007a}
\bibinfo{author}{\bibfnamefont{T.}~\bibnamefont{M{\"{u}}ller}},
  \bibinfo{author}{\bibfnamefont{S.}~\bibnamefont{F{\"{o}}lling}},
  \bibinfo{author}{\bibfnamefont{A.}~\bibnamefont{Widera}}, \bibnamefont{and}
  \bibinfo{author}{\bibfnamefont{I.}~\bibnamefont{Bloch}},
  \bibinfo{journal}{Phys. Rev. Lett.} \textbf{\bibinfo{volume}{99}},
  \bibinfo{pages}{200405} (\bibinfo{year}{2007}).

\bibitem[{\citenamefont{Isacsson and Girvin}(2005)}]{Isacsson2005a}
\bibinfo{author}{\bibfnamefont{A.}~\bibnamefont{Isacsson}} \bibnamefont{and}
  \bibinfo{author}{\bibfnamefont{S.~M.} \bibnamefont{Girvin}},
  \bibinfo{journal}{Phys. Rev. A} \textbf{\bibinfo{volume}{72}},
  \bibinfo{pages}{053604} (\bibinfo{year}{2005}).

\bibitem[{\citenamefont{Scarola and Sarma}(2005)}]{Scarola2005a}
\bibinfo{author}{\bibfnamefont{V.~W.} \bibnamefont{Scarola}} \bibnamefont{and}
  \bibinfo{author}{\bibfnamefont{S.~D.} \bibnamefont{Sarma}},
  \bibinfo{journal}{Phys. Rev. Lett.} \textbf{\bibinfo{volume}{95}},
  \bibinfo{pages}{033003} (\bibinfo{year}{2005}).

\bibitem[{\citenamefont{Liu and Wu}(2006)}]{Liu2006a}
\bibinfo{author}{\bibfnamefont{W.~V.} \bibnamefont{Liu}} \bibnamefont{and}
  \bibinfo{author}{\bibfnamefont{C.}~\bibnamefont{Wu}}, \bibinfo{journal}{Phys.
  Rev. A} \textbf{\bibinfo{volume}{74}}, \bibinfo{pages}{013607}
  (\bibinfo{year}{2006}).

\bibitem[{\citenamefont{Xu and Fisher}(2007)}]{Xu2007a}
\bibinfo{author}{\bibfnamefont{C.}~\bibnamefont{Xu}} \bibnamefont{and}
  \bibinfo{author}{\bibfnamefont{M.~P.~A.} \bibnamefont{Fisher}},
  \bibinfo{journal}{Phys. Rev. B} \textbf{\bibinfo{volume}{75}},
  \bibinfo{pages}{104428} (\bibinfo{year}{2007}).

\bibitem[{\citenamefont{Wu and Zhai}(2008)}]{Wu2008a}
\bibinfo{author}{\bibfnamefont{K.}~\bibnamefont{Wu}} \bibnamefont{and}
  \bibinfo{author}{\bibfnamefont{H.}~\bibnamefont{Zhai}},
  \bibinfo{journal}{Phys. Rev. B} \textbf{\bibinfo{volume}{77}},
  \bibinfo{pages}{174431} (\bibinfo{year}{2008}).

\bibitem[{\citenamefont{Wu and Sarma}(2008)}]{CongjunWu2008a}
\bibinfo{author}{\bibfnamefont{C.}~\bibnamefont{Wu}} \bibnamefont{and}
  \bibinfo{author}{\bibfnamefont{S.~D.} \bibnamefont{Sarma}},
  \bibinfo{journal}{Phys. Rev. B} \textbf{\bibinfo{volume}{77}},
  \bibinfo{pages}{235107} (\bibinfo{year}{2008}).

\bibitem[{\citenamefont{Martikainen et~al.}(2008)\citenamefont{Martikainen,
  Lundh, and Paananen}}]{Martikainen2008a}
\bibinfo{author}{\bibfnamefont{J.-P.} \bibnamefont{Martikainen}},
  \bibinfo{author}{\bibfnamefont{E.}~\bibnamefont{Lundh}}, \bibnamefont{and}
  \bibinfo{author}{\bibfnamefont{T.}~\bibnamefont{Paananen}},
  \bibinfo{journal}{Phys. Rev. A} \textbf{\bibinfo{volume}{78}},
  \bibinfo{pages}{023607} (\bibinfo{year}{2008}).

\bibitem[{\citenamefont{K{\"{o}}hl et~al.}(2005)\citenamefont{K{\"{o}}hl,
  Moritz, St{\"{o}}ferle, G{\"{u}}nter, and Esslinger}}]{Kohl2005a}
\bibinfo{author}{\bibfnamefont{M.}~\bibnamefont{K{\"{o}}hl}},
  \bibinfo{author}{\bibfnamefont{H.}~\bibnamefont{Moritz}},
  \bibinfo{author}{\bibfnamefont{T.}~\bibnamefont{St{\"{o}}ferle}},
  \bibinfo{author}{\bibfnamefont{K.}~\bibnamefont{G{\"{u}}nter}},
  \bibnamefont{and}
  \bibinfo{author}{\bibfnamefont{T.}~\bibnamefont{Esslinger}},
  \bibinfo{journal}{Phys. Rev. Lett.} \textbf{\bibinfo{volume}{94}},
  \bibinfo{pages}{080403} (\bibinfo{year}{2005}).

\bibitem[{\citenamefont{Baillie and Blakie}(2009)}]{Baillie2009a}
\bibinfo{author}{\bibfnamefont{D.}~\bibnamefont{Baillie}} \bibnamefont{and}
  \bibinfo{author}{\bibfnamefont{P.~B.} \bibnamefont{Blakie}}
  (\bibinfo{year}{2009}), \eprint{arXiv:0906.4606}.

\bibitem[{\citenamefont{Georges}(2007)}]{Georges2007a}
\bibinfo{author}{\bibfnamefont{A.}~\bibnamefont{Georges}}
  (\bibinfo{year}{2007}), \eprint{arXiv:cond-mat/0702122}.

\bibitem[{\citenamefont{Law et~al.}(1998)\citenamefont{Law, Pu, and
  Bigelow}}]{Law1998a}
\bibinfo{author}{\bibfnamefont{C.~K.} \bibnamefont{Law}},
  \bibinfo{author}{\bibfnamefont{H.}~\bibnamefont{Pu}}, \bibnamefont{and}
  \bibinfo{author}{\bibfnamefont{N.~P.} \bibnamefont{Bigelow}},
  \bibinfo{journal}{Phys. Rev. Lett.} \textbf{\bibinfo{volume}{81}},
  \bibinfo{pages}{5257} (\bibinfo{year}{1998}).

\bibitem[{\citenamefont{Stamper-Kurn et~al.}(1998)\citenamefont{Stamper-Kurn,
  Andrews, Chikkatur, Inouye, Miesner, Stenger, and
  Ketterle}}]{Stamper-Kurn1998b}
\bibinfo{author}{\bibfnamefont{D.~M.} \bibnamefont{Stamper-Kurn}},
  \bibinfo{author}{\bibfnamefont{M.~R.} \bibnamefont{Andrews}},
  \bibinfo{author}{\bibfnamefont{A.~P.} \bibnamefont{Chikkatur}},
  \bibinfo{author}{\bibfnamefont{S.}~\bibnamefont{Inouye}},
  \bibinfo{author}{\bibfnamefont{H.-J.} \bibnamefont{Miesner}},
  \bibinfo{author}{\bibfnamefont{J.}~\bibnamefont{Stenger}}, \bibnamefont{and}
  \bibinfo{author}{\bibfnamefont{W.}~\bibnamefont{Ketterle}},
  \bibinfo{journal}{Phys. Rev. Lett.} \textbf{\bibinfo{volume}{80}},
  \bibinfo{pages}{2027} (\bibinfo{year}{1998}).

\bibitem[{\citenamefont{Duan}(2008)}]{Duan2008a}
\bibinfo{author}{\bibfnamefont{L.-M.} \bibnamefont{Duan}},
  \bibinfo{journal}{Euro. Phys. Lett.} \textbf{\bibinfo{volume}{81}},
  \bibinfo{pages}{20001} (\bibinfo{year}{2008}).

\bibitem[{\citenamefont{Li et~al.}(2006)\citenamefont{Li, Yu, Dudarev, and
  Niu}}]{Li2006a}
\bibinfo{author}{\bibfnamefont{J.}~\bibnamefont{Li}},
  \bibinfo{author}{\bibfnamefont{Y.}~\bibnamefont{Yu}},
  \bibinfo{author}{\bibfnamefont{A.~M.} \bibnamefont{Dudarev}},
  \bibnamefont{and} \bibinfo{author}{\bibfnamefont{Q.}~\bibnamefont{Niu}},
  \bibinfo{journal}{New J. Phys.} \textbf{\bibinfo{volume}{8}},
  \bibinfo{pages}{154} (\bibinfo{year}{2006}).

\bibitem[{\citenamefont{Buonsante et~al.}(2008)\citenamefont{Buonsante,
  Giampaolo, Illuminati, Penna, and Vezzani}}]{Buonsante2008a}
\bibinfo{author}{\bibfnamefont{P.}~\bibnamefont{Buonsante}},
  \bibinfo{author}{\bibfnamefont{S.}~\bibnamefont{Giampaolo}},
  \bibinfo{author}{\bibfnamefont{F.}~\bibnamefont{Illuminati}},
  \bibinfo{author}{\bibfnamefont{V.}~\bibnamefont{Penna}}, \bibnamefont{and}
  \bibinfo{author}{\bibfnamefont{A.}~\bibnamefont{Vezzani}},
  \bibinfo{journal}{Phys. Rev. Lett.} \textbf{\bibinfo{volume}{100}},
  \bibinfo{pages}{240402} (\bibinfo{year}{2008}).

\bibitem[{\citenamefont{Larson et~al.}(2009)\citenamefont{Larson, Collin, and
  Martikainen}}]{Larson2009a}
\bibinfo{author}{\bibfnamefont{J.}~\bibnamefont{Larson}},
  \bibinfo{author}{\bibfnamefont{A.}~\bibnamefont{Collin}}, \bibnamefont{and}
  \bibinfo{author}{\bibfnamefont{J.-P.} \bibnamefont{Martikainen}},
  \bibinfo{journal}{Phys. Rev. A} \textbf{\bibinfo{volume}{79}},
  \bibinfo{pages}{033603} (\bibinfo{year}{2009}).

\bibitem[{\citenamefont{Watanabe and Pethick}(2007)}]{Watanabe2007}
\bibinfo{author}{\bibfnamefont{G.}~\bibnamefont{Watanabe}} \bibnamefont{and}
  \bibinfo{author}{\bibfnamefont{C.~J.} \bibnamefont{Pethick}},
  \bibinfo{journal}{Phys. Rev. A} \textbf{\bibinfo{volume}{76}},
  \bibinfo{pages}{021605(R)} (\bibinfo{year}{2007}).

\bibitem[{\citenamefont{Lim et~al.}(2008)\citenamefont{Lim, Smith, and
  Hemmerich}}]{Lim2008a}
\bibinfo{author}{\bibfnamefont{L.-K.} \bibnamefont{Lim}},
  \bibinfo{author}{\bibfnamefont{C.~M.} \bibnamefont{Smith}}, \bibnamefont{and}
  \bibinfo{author}{\bibfnamefont{A.}~\bibnamefont{Hemmerich}},
  \bibinfo{journal}{Phys. Rev. Lett.} \textbf{\bibinfo{volume}{100}},
  \bibinfo{pages}{130402} (\bibinfo{year}{2008}).

\bibitem[{\citenamefont{Leonhardt and Volovik}(2000)}]{Leonhardt2000a}
\bibinfo{author}{\bibfnamefont{U.}~\bibnamefont{Leonhardt}} \bibnamefont{and}
  \bibinfo{author}{\bibfnamefont{G.~E.} \bibnamefont{Volovik}},
  \bibinfo{journal}{JETP Lett.} \textbf{\bibinfo{volume}{72}},
  \bibinfo{pages}{46} (\bibinfo{year}{2000}).

\bibitem[{\citenamefont{Martikainen et~al.}(2002)\citenamefont{Martikainen,
  Collin, and Suominen}}]{Martikainen2002b}
\bibinfo{author}{\bibfnamefont{J.~P.} \bibnamefont{Martikainen}},
  \bibinfo{author}{\bibfnamefont{A.}~\bibnamefont{Collin}}, \bibnamefont{and}
  \bibinfo{author}{\bibfnamefont{K.~A.} \bibnamefont{Suominen}},
  \bibinfo{journal}{Phys. Rev. A} \textbf{\bibinfo{volume}{66}},
  \bibinfo{pages}{053604} (\bibinfo{year}{2002}).

\bibitem[{\citenamefont{Machholm et~al.}(2003)\citenamefont{Machholm, Pethick,
  and Smith}}]{Machholm2003a}
\bibinfo{author}{\bibfnamefont{M.}~\bibnamefont{Machholm}},
  \bibinfo{author}{\bibfnamefont{C.~J.} \bibnamefont{Pethick}},
  \bibnamefont{and} \bibinfo{author}{\bibfnamefont{H.}~\bibnamefont{Smith}},
  \bibinfo{journal}{Phys. Rev. A} \textbf{\bibinfo{volume}{67}},
  \bibinfo{pages}{053613} (\bibinfo{year}{2003}).

\bibitem[{\citenamefont{Machholm et~al.}(2004)\citenamefont{Machholm, Nicolin,
  Pethick, and Smith}}]{Machholm2004a}
\bibinfo{author}{\bibfnamefont{M.}~\bibnamefont{Machholm}},
  \bibinfo{author}{\bibfnamefont{A.}~\bibnamefont{Nicolin}},
  \bibinfo{author}{\bibfnamefont{C.~J.} \bibnamefont{Pethick}},
  \bibnamefont{and} \bibinfo{author}{\bibfnamefont{H.}~\bibnamefont{Smith}},
  \bibinfo{journal}{Phys. Rev. A} \textbf{\bibinfo{volume}{69}},
  \bibinfo{pages}{043604} (\bibinfo{year}{2004}).

\end{thebibliography}
\end{document}